\documentclass[preprint]{sigchi}

\usepackage{balance}       % to better equalize the last page
\usepackage{graphics}      % for EPS, load graphicx instead 
\usepackage[T1]{fontenc}   % for umlauts and other diaeresis
\usepackage{txfonts}
\usepackage{mathptmx}

\usepackage{color}
\usepackage{booktabs}
\usepackage{textcomp}
\usepackage{lipsum}
\usepackage{subcaption}
\usepackage{multirow}

\usepackage{microtype}        % Improved Tracking and Kerning
\usepackage{ccicons}          % Cite your images correctly!
\def\plaintitle{Enhanced Touchable Projector-depth System with Deep Hand Pose Estimation}

\def\emptyauthor{Zhi Chai, Roy Shilkrot}
\def\plainkeywords{Projection Surface; Touch Detection; Hand Pose Estimation}

\makeatletter
\def\url@leostyle{%
  \@ifundefined{selectfont}{
    \def\UrlFont{\sf}
  }{
    \def\UrlFont{\small\bf\ttfamily}
  }}
\makeatother
\def\pprw{8.5in}
\def\pprh{11in}

\usepackage[pdflang={en-US},pdftex]{hyperref}

\definecolor{linkColor}{RGB}{6,125,233}
\hypersetup{%
  pdftitle={\plaintitle},
  pdfauthor={\emptyauthor},
  pdfkeywords={\plainkeywords},
  pdfdisplaydoctitle=true, % For Accessibility
  bookmarksnumbered,
  pdfstartview={FitH},
  colorlinks,
  citecolor=black,
  filecolor=black,
  linkcolor=black,
  urlcolor=linkColor,
  breaklinks=true,
  hypertexnames=false
}

\makeatletter
\def\@copyrightspace{\relax}
\makeatother

\begin{document}

\title{\plaintitle}

\numberofauthors{2}
\author{%
  \alignauthor{Zhi Chai\\
    \affaddr{Stony Brook University}\\
    \affaddr{Computer Science Department}\\
    \affaddr{Stony Brook, NY 11794, USA}\\
    \email{zhchai@cs.stonybrook.edu}}\\
  \alignauthor{Roy Shilkrot\\
    \affaddr{Stony Brook University}\\
	\affaddr{Computer Science Department}\\
    \affaddr{Stony Brook, NY 11794, USA}\\
    \email{roys@cs.stonybrook.edu}}\\
}

\maketitle

\begin{abstract}
Touchable projection with structured light range cameras is a prolific medium for large interaction surfaces, affording multiple simultaneous users and simple, cheap setup. However robust touch detection in such projector-depth systems is difficult to achieve due to measurement noise. We propose a novel combination of surface touch detection and a deep network for hand pose estimation, which aids in detecting both on- and above-surface hand gestures, disambiguating multiple touch fingers, as well as recovering fingertip positions in face of noisy input. We present the details of our GPU-accelerated system and an evaluation of its performance, as well as applications such as an enhanced virtual keyboard that utilizes the added features.
\end{abstract}


\section{Introduction}
Interactive projection surfaces are becoming a regular method for augmenting large surfaces such as tabletops, walls and floors. Recently, wearable augmented and virtual reality technologies receive a thrust of attention, however their meager visual experience created a renewed opportunity for the more visually natural augmentation-by-projection. Projection technology has matured and range cameras have commodified making non-instrumented touch-enabled surfaces proliferate in public spaces such as retail stores, office space and industrial settings. The favorable environment for touchable projection has boosted academic research, particularly after the public release of the Microsoft Kinect (and its derivatives) \cite{Wilson:2010}, as well as commercial ventures~\cite{ubiinteractive}. Our system explores how new techniques for real-time interaction can add even more capabilities to these existing methods.

\begin{figure}[h!]
\includegraphics[width=\linewidth]{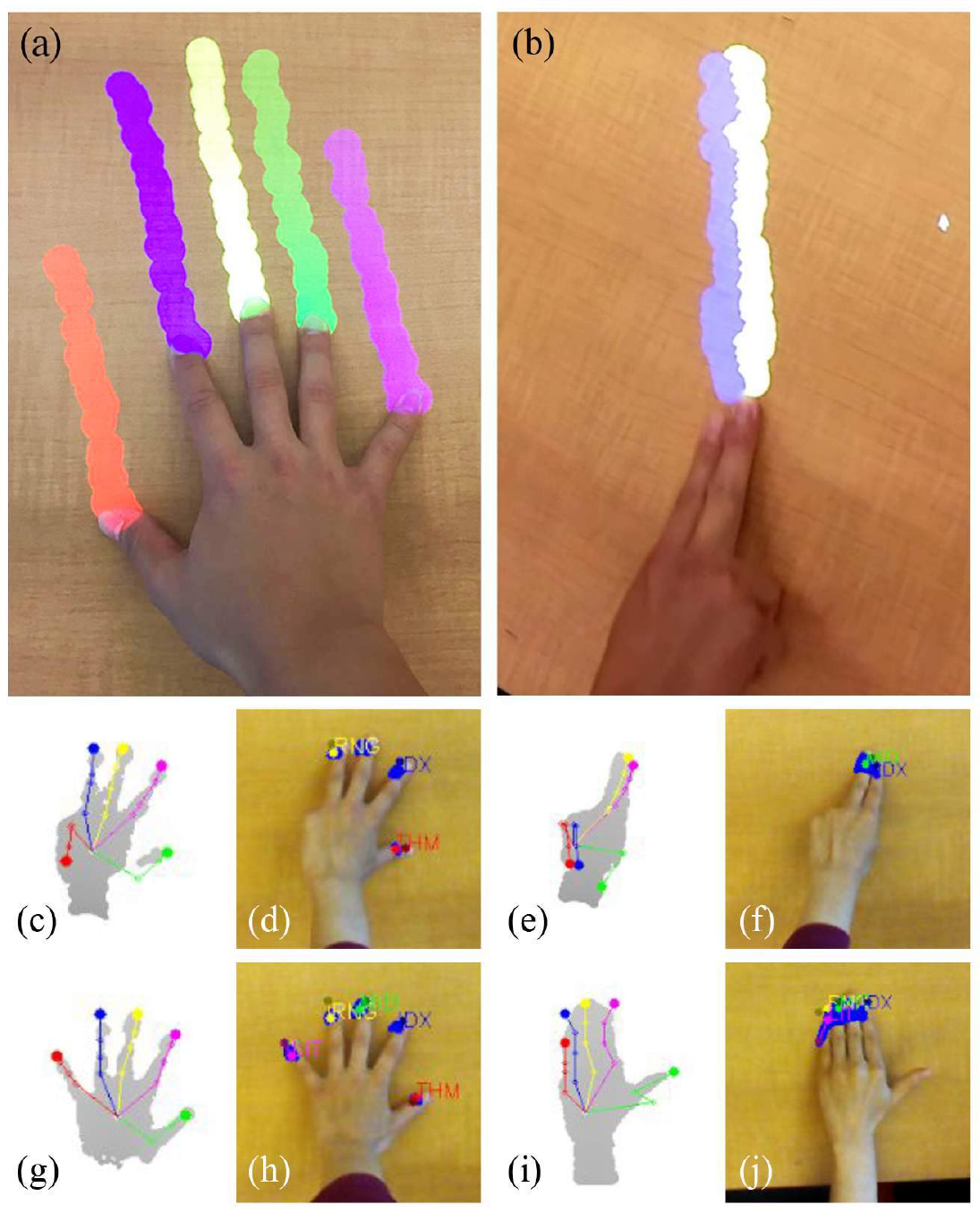}
\caption{Results of our system. (a, b) projected Finger designation (note the color of the index and middle fingers). (d, f, h, j) decomposed touch-blobs (colored blue). (c, e, g, i) estimated pose of 20 hand joints.}
\label{fig:teaser}
\end{figure}

The unprecedented explosion of highly parametric but real-time techniques for segmentation and pose estimation, since the breakthrough in efficient training of deep convolutional neural networks (CNNs), evoked a strong interest in articulated hand pose estimation. With proven methods ready at hand, the application to surface touch seems straightforward, however there is not an immediate transfer, due to the extreme proximity of the hand to a surface. The overwhelming majority of depth-based hand posers focus on in-air gestures rather than surface touch, therefore the approach as well as the training datasets must be adapted.

\begin{figure*}[h!]
\centering
\includegraphics[width=0.8\linewidth]{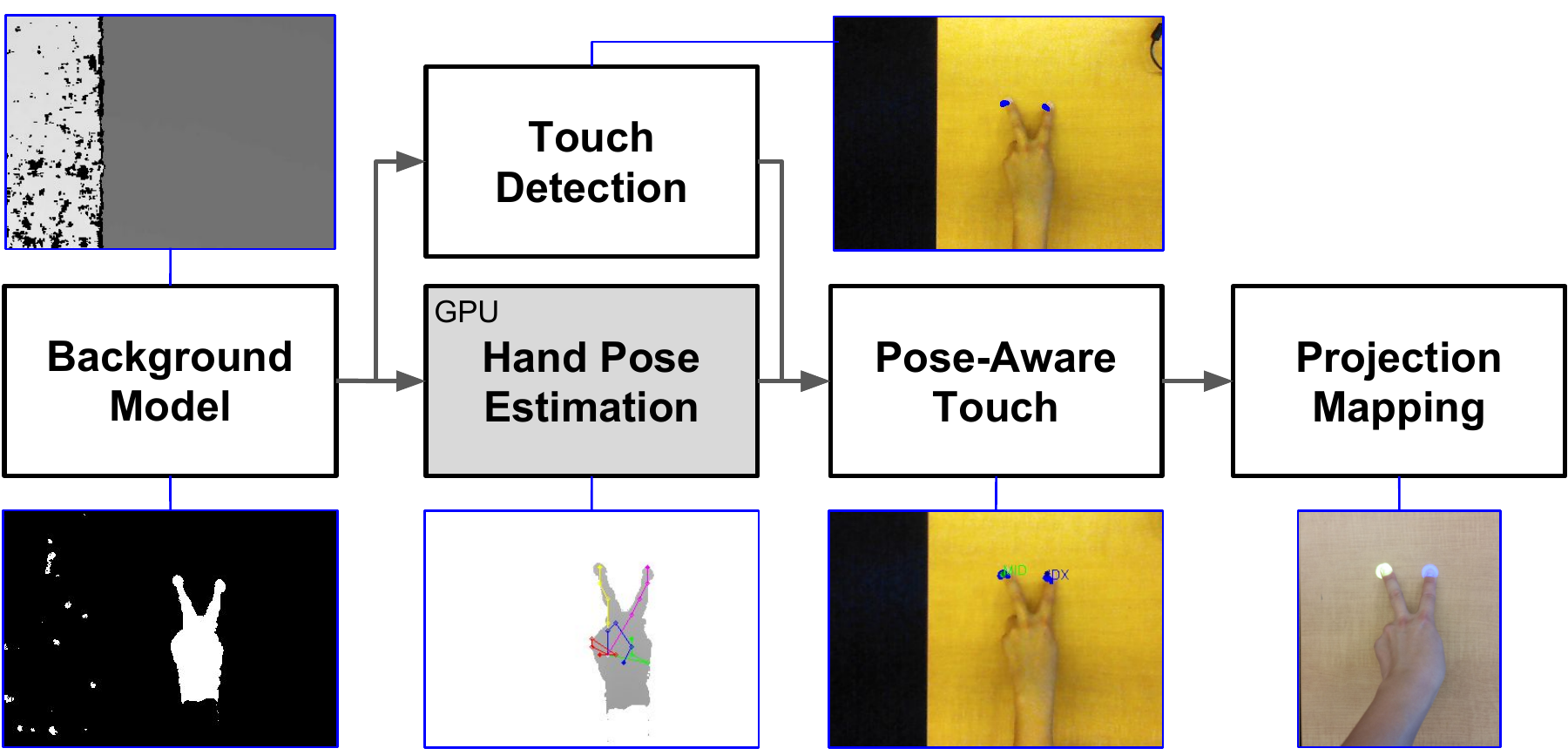}
\caption{Processing pipeline: Background modeling, segmentation, hand pose estimation, and pose-aware touch detection.}
\label{fig:overview}
\end{figure*}

\textbf{Contributions of this work.} We developed a deep CNN-based hand pose estimation directly trained on top-viewpoint surface touch operations, which resulted in numerous opportunities to augment existing touchable projection (partially illustrated in fig.~\ref{fig:teaser}). First, with fully articulated pose estimation, finger identification is available and can enrich multi-touch information. Second, while current touch detection methods are primarily based on analysis of near-surface (touch-) blobs, our proposed method is derived from the pose of the hand and therefore can decompose blobs of multiple fingers pressed together when touching the surface (see fig.~\ref{fig:teaser} (b), (f), (j)). Finally, our deep CNN, trained over a large dataset of touch samples, provides predicted fingertip locations even in presence of occlusion or measurement noise (see  fig.~\ref{fig:teaser} (c)), a common occurrence in range imaging due to stereo disparity, which is highly disruptive for connected-component analysis methods. 

In summary, this work contributes a novel pose-aware touch detection in the following aspects:

\begin{itemize}
  \item Touch finger designation and disambiguation
  \item Hand gesture detection
  \item On and above surface detection
  \item Robust to sensor noise
  \item A top view hand pose dataset 
\end{itemize}

We hereby describe our findings in hand pose estimation for enhanced touchable projection, starting with a literature review, continuing with the system description and concluding with evaluations.

\section{Related Work}
Since the early interactive table-top surfaces of Wellner~\cite{wellner1993interacting} and Ullmer~\cite{ullmer2000emerging}, much and more has happened. Implementations of touchable projection using cameras suggested intuitive interaction~\cite{wilson2005playanywhere} even before touch-screens were widely available, as well as novel sensing capabilities such as 3D scanning and augmentation-by-projection~\cite{raskar2006ilamps}. Our latest contribution to this endeavor combines deep hand pose estimation and multi-touch detection mechanics from range imaging, which we review in the following section.

\subsection{Surface Multi-Touch Detection}
More than a few works exist for identifying surface touch positions from depth cameras, so in interest of brevity we hereby recount only the most relevant, recent ones. Wilson et al. presented a method to use a single depth camera to retrieve touch information~\cite{Wilson:2010}, by way of a depth-background model and examining the relative depth to the background to identify touch points under finger thickness constraints. Murugappan et al.~\cite{Murugappan:2012} extended~\cite{Wilson:2010} in identifying touch points by doing connected components analysis and discarding blobs based on area constraints, as well as introduced a method for touch-gesture detection based on the blobs. Xiao et al.~\cite{Xiao:2016} go beyond the depth image to perform edge detection on the sensor's infrared (IR) stream to get hand boundaries and fingertips, then pruning touching fingertips by examining a small depth pixel neighborhood. Cadena et al.~\cite{Cadena:2016} use depth image to first segment the arm and hand using k-means, then fingertips are detected using geometric analysis and refined based on IR edge detection as in~\cite{Xiao:2016}.
In our work, we initially follow~\cite{Wilson:2010,Murugappan:2012} but vastly extend these with data-driven segmentation and hand pose estimation, instead of a parametric approach.

\subsection{Hand Pose Estimation}

Data-driven methods ~\cite{SupancicRYSR15,Riegler2015} have had a great success recently in hand pose recovery. Tang et al.~\cite{Tang14} use Regression Forest to regress the location of joints directly in 3D space, whereas Tompson et al.~\cite{Tompson:2014} regress the location of joints in 2D heatmaps by using a deep neural network. Inspired by~\cite{Tompson:2014}, Ge et al.~\cite{GeLYT16} train several networks on 3 projected planes to improve predictions under heavy occlusions. Rather than use a per-joint estimation, Sun et al.~\cite{Sun15} simply regress in a holistic way with hierarchical structure.

We generally follow the work of ~\cite{Tompson:2014}, however unlike our setup it is targeted at air-gesture detection as well as reliant on a particular training data collection scheme. To that end we adapted \cite{Riegler2015} to top-view imaging and extended it to create a much more realistic rendering.

\section{Our System}
Our touchable projection system is based on a combination of methods: Background modeling using depth-pixel histograms, arm-hand segmentation using background subtraction, hand pose estimation using a CNN, and finally pose-aware touch detection utilizing the skeleton information visualized by a calibrated projector-camera setup. See Fig.\ref{fig:overview} for the overview of our system.

\subsection{Surface Modeling and Segmentation}
Following~\cite{Wilson:2010} we model the surface to be able to detect touch on it. We use the first 30 frames towards building the background model while the surface is empty. Thereafter we calculate a per-pixel depth-difference histogram w.r.t a reference frame, disregarding all invalid depth values, and then determine the background model from this histogram using the same policy as in \cite{Wilson:2010}. We calculate the valid depth range based on the background model using an above-surface range of 2mm for touch, and anything else above the table is considered for finding arms and hands. 

From the arm and hand we would like to extract just the hand part to feed into the CNN poser. To that end we localize the hand region by thresholding pixels in range as in \cite{Murugappan:2012}.

\begin{figure}[h!]
\centering
\includegraphics[width=\linewidth, scale = 0.7]{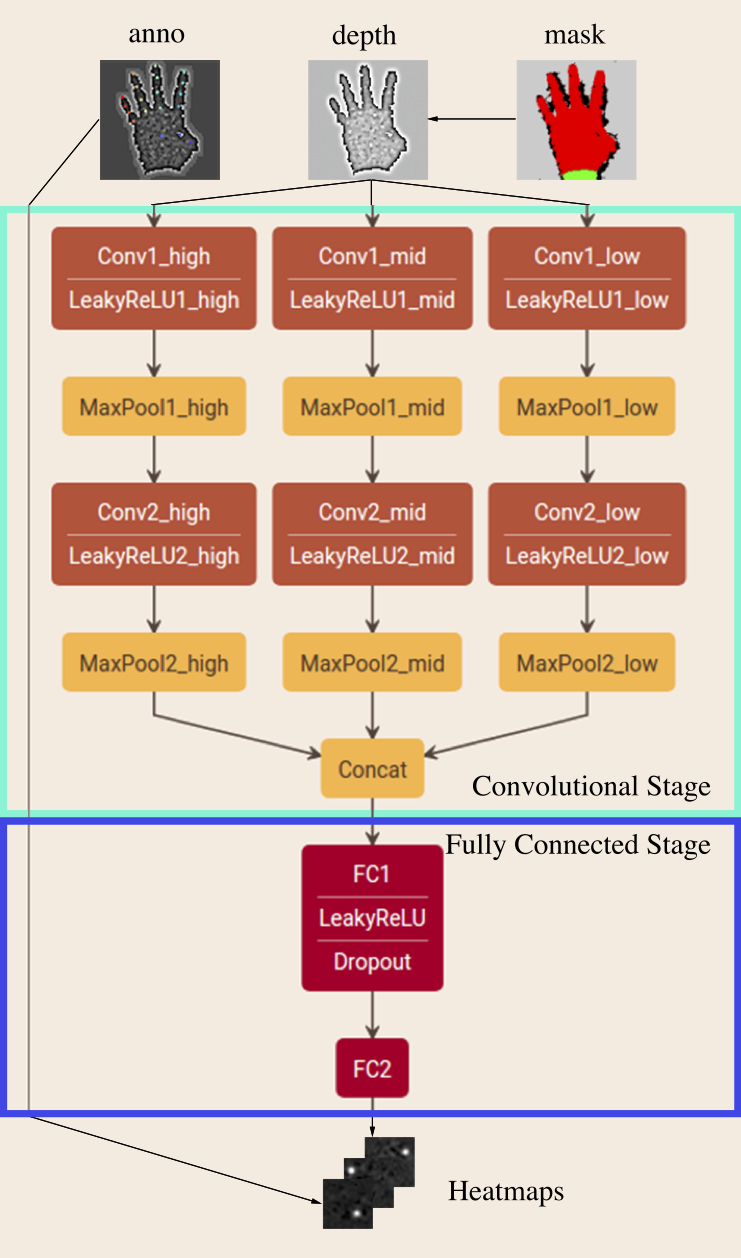}
\caption{The topology of our deep CNN for hand pose estimation.}
\label{fig:deepcnn}
\end{figure}

\subsection{Top View Hand Dataset}
\begin{figure*}[h!]
\centering
\includegraphics[width=\linewidth]{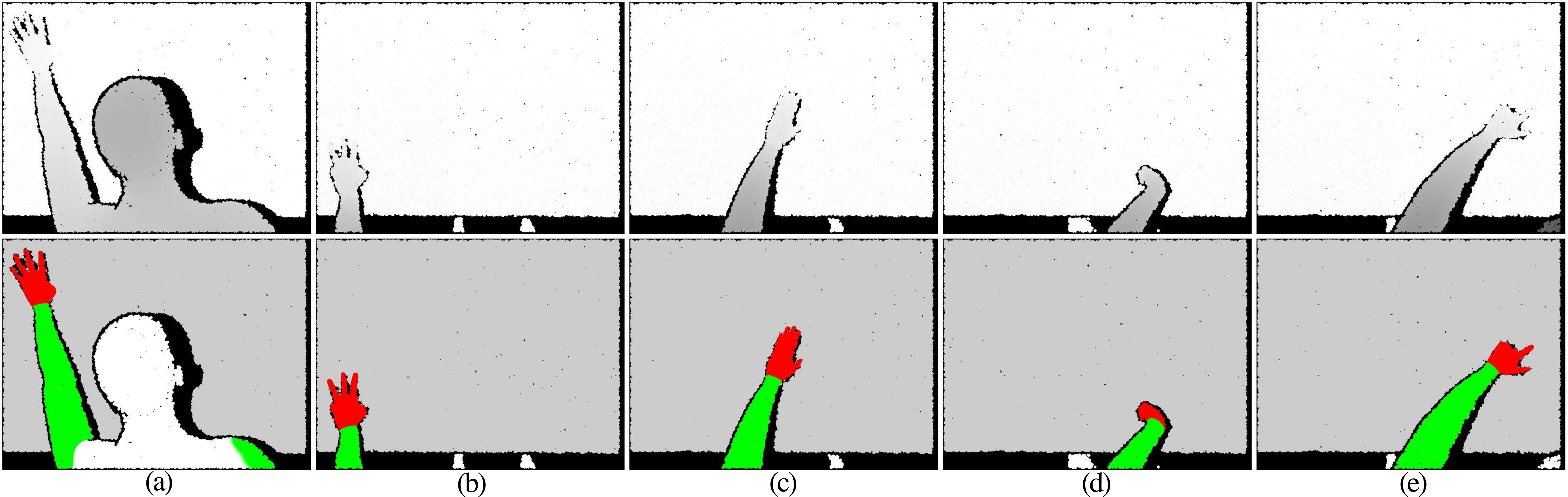}
\caption{A sample of various touch poses in our dataset. (a)-(e) Top row: simulated range images. Bottom row: corresponding RGB images with hand segmentation. In total our dataset contains over 20,000 samples.}
\label{fig:dataset}
\end{figure*}

Widely used hand pose datasets like NYU~\cite{Tompson:2014}, ICL~\cite{Tang14} and UCI-EGO~\cite{RogezSKMR14} focus on in-air hand pose detection, while in our method we track hands in close surface proximity. Thus we design a multifarious dataset of surface-touching as well as in-air hand poses from a top view, based on the work of~\cite{Riegler2015} that simulates human hands in a 3D virtual environment. Synthetic data, as used in ~\cite{SupancicRYSR15}, has a drawback in which the generated virtual images are usually error free, which is not the case in real data from a range sensor. To overcome this problem and make our data more robust to noise, we utilize Blensor~\cite{Gschwandtner11}, a Blender add-on that mimics real distortion effects in a structured-light camera by deliberately introducing errors. See Fig.\ref{fig:dataset} for a sample of images from our synthetic dataset, and note the realism effects on the range image that Blensor introduces.

In order to create a large and representative hand pose dataset for both on and above surface hand detection, we manually designed 7, 10, 11 and 14 poses for deltoid, upper arm, forearm and finger joints respectively. The combination of all poses of the 4 parts gives us 10780 rendered images. We then horizontally flip all the images to simulate the left hand. In general, we have more than 20k images to represent ordinary in-air and touch poses from top view. The corresponding RGB images help to localize the hand region in the training phase, and an annotation of each joint on the hand model is generated automatically during the rendering process.

\subsection{Deep Hand Pose Estimation}
All the depth images from our top-view dataset are cropped, normalized to the [-1, 1] range according to the corresponding hand-region mask images, and Local Contrast Normalization (LCN)~\cite{JarrettKRL09} is applied to emphasize geometric discontinuities. The resulting depth images are finally fed into our neural network as displayed in Fig.~\ref{fig:deepcnn}.

Following~\cite{Tompson:2014}, we use a two-stage neural network trained on our synthetic dataset. For prediction, we either look for just the 5 fingertips or the full 20-joints hand skeleton, spatially embedded in a heatmap for each joint or fingertip. 
The network is implemented using Caffe~\cite{jia2014caffe} with GPU support. We employ multi-resolution feature detectors to improve the results. In each feature detector, two convolution layers followed by leaky ReLU and max pooling layers are used to extract abstract features. Outputs from three banks are merged together and then feed into another neural network consisting of two fully connected layers. Dropout with probability of 0.7 is also applied throughout. See fig.~\ref{fig:deepcnn} for a visualization of this topology. 
Adding a Gaussian fitting on the heatmap output of the two-stage network yields the most confident position of each joint. 

By knowing the 2D position of each hand joint, we can infer the 2.5D position by looking up the depth value in the depth image, however in our work 2D information is sufficient. Similar to \cite{Tompson:2014}, our fitness function is based on the angles formed by the estimated joints positions. For regularization, we add an exponential penalty to any angle that is outside of the valid range.

\subsection{Pose-aware Touch Detection}
We employ connected components analysis to determine the contours of fingers touching the surface. We devise a similar method to~\cite{Wilson:2010,Murugappan:2012} to segment the hand, setting the valid depth slice to be between 2 and 170 mm, whereas under 2 mm we consider the be in touch range. To remove noise we apply box filtering, blob-area thresholding, and use the negative bounding rectangle of the blob to discard any noisy touch points outside of it. Similar to~\cite{Xiao:2016} we detect and discard hover points.

At this point we have the touch locations in addition to the estimated fingertips from the hand pose estimation. We move on to mapping the fingertips to the touch blobs to get finger designation (assign a finger label to each blob). For mapping we utilize the Hungarian matching scheme, using $\textrm{L}_2$ distances for the cost matrix. To this optimization we add temporal information from the last frame to increase consistency in tracking:
\[
\textrm{Cost}(f,t) = 
\begin{cases}
\lVert f-t \rVert + \lVert f' - t\rVert & \textrm{if } \lVert f-t \rVert < D_t \\
D_e & \textrm{otherwise}
\end{cases}
\]
where $f$ is the finger point, $t$ is the touch point and $f'$ is the finger point in the last frame. $D_t$ is a threshold for considering a finger-touch pair, if crossed a very big penalty is used for the cost $D_e$. At the end of each frame we assign $f' := f$.

Moreover, a touch blob can span multiple fingers if they are close or pressed together in the hand pose. Using the fingers tips from the pose estimation we can split that blob into the underlying fingers, see Fig.\ref{fig:teaser}. This is done by uniformly sampling the bounding rectangle of the blob at a rate of 9 pixels in both x and y axes, s.t. the sampled points lie within the touch blob. We treat the sampled points as touch locations and match with the nearest fingertip. If a blob is too small we treat it as a single touch point that is the center of the blob. Combining the touch estimates and fingertip estimates allows us to obtain more detailed information about touches and the user's gesture.

\begin{figure}[h!]
\centering
\includegraphics[width=\linewidth, scale = 0.8]{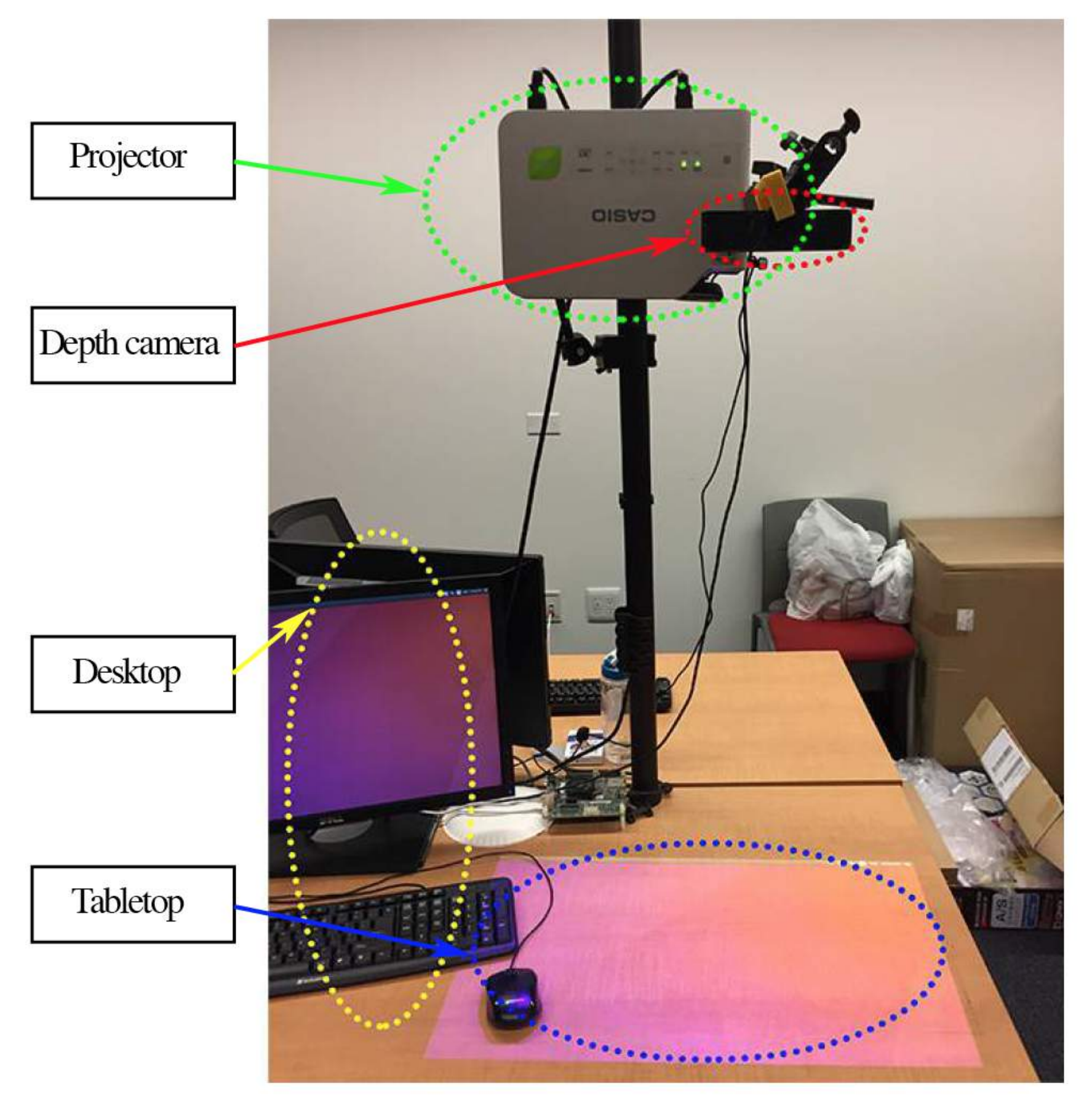}
\caption{The physical setup of our system.}
\label{fig:setup}
\end{figure}

\section{Implementation Details}
The system is implemented using C++, OpenCV, Caffe~\cite{jia2014caffe} and CUDA in a Ubuntu environment. We used an Nvidia Titan Xp GPU, an Intel i7-5930K CPU, and a system with 32Gb RAM. All depth frames (640$\times$480 pixels) are taken by the structured-light sensor named Astra (distributed by Orbbec). 
The GPU memory footprint is 1Gb (alloted to CNN and working buffers), the application takes up ~2Gb of the system RAM.

\subsection{Deep Neural Network and Above-surface Tracking}
We train the network on our own version of Caffe~\cite{jia2014caffe} with GPU enabled. The network is trained on a dataset with the size of 20k, we use the batch size as 64, the learning rate as 0.00004, the momentum as 0.9 and the weight decay as 0.0005. The training is stopped after 100 epochs to prevent overfitting. 

The results of the hand pose estimation can be found in Fig.\ref{fig:teaser}(c, e, g, i). Our prediction is robust to sensor noise as presented in Fig.\ref{fig:teaser}(c), the thumb can even be successfully predicted when it is detached from the hand blob.

Due to the employment of hand pose estimation, we can also detect above-surface hand poses which might be used to augment the touch detection.

\section{Evaluation}
We designed four experiments to measure our on-surface touch tracking ability following \cite{Xiao:2016}. Since we are using a structured light (S-L) sensor in a Linux environment, we were unable to compare our results directly to the recent method in \cite{Xiao:2016}, which employs a Kinect 2 time-of-flight (ToF) sensor in a Windows environment. S-L sensors such as we use cannot produce a clean IR frame as ToF sensors do, since they use a laser projector that adds a strong pattern to the scene image (to calculate the range), preventing from performing reliable edges analysis. 
We therefore compare our results to the canonical Wilson et al.~\cite{Wilson:2010} method. 

The four interaction evaluation tasks are: $1)$ touch the center of a projected cross 20 times; $2)$ track a horizontal line; $3)$ track a vertical line; $4)$ track a circle. A total of 10 subjects participate in the experiments and Fig.\ref{fig:setup} serves as an illustration of the working environment.

\begin{figure}[h!]
\centering
\includegraphics[width=\linewidth]{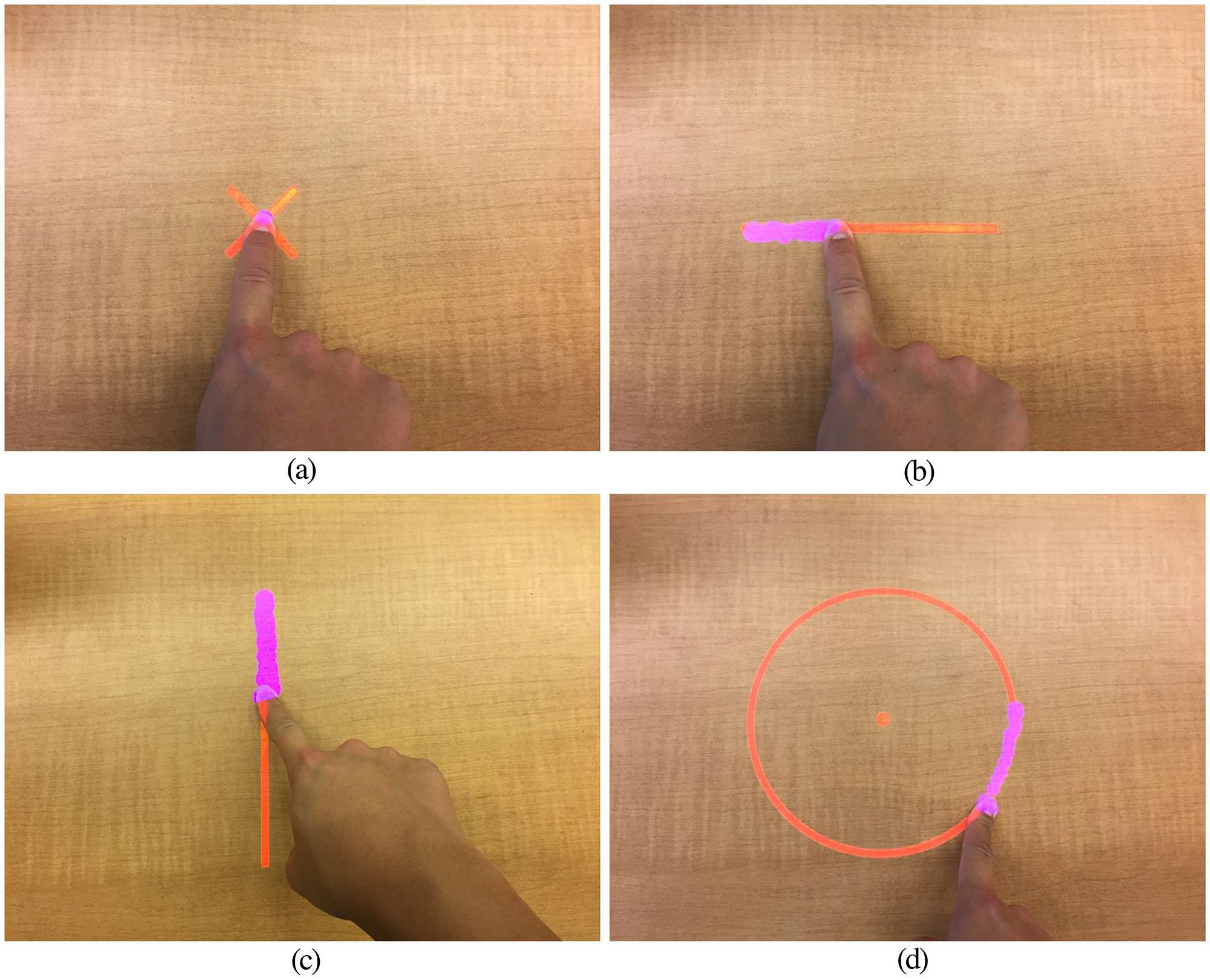}
\caption{Four tasks to evaluate the accuracy of our method. (a) touch the center of a cross. (b) track the horizontal line on the desk. (c) track the vertical line on the desk. (d) track the circle on the desk. Error bars are Standard Error.}
\label{fig:tasks}
\end{figure}

\begin{figure*}
    \centering
 \begin{tabular}{ccc}
   \includegraphics[width=.45\textwidth]{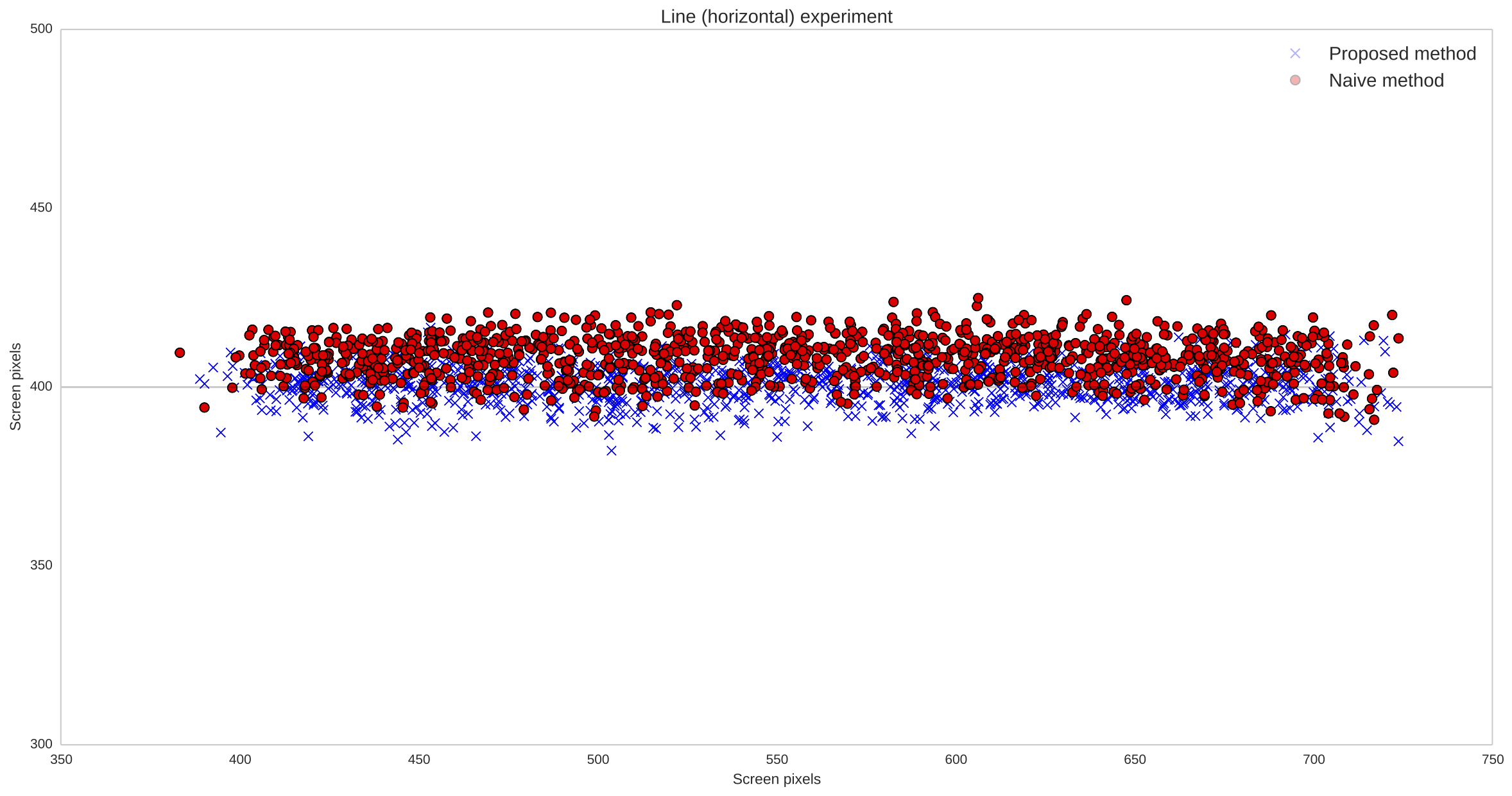} &%
   \includegraphics[width=.25\textwidth]{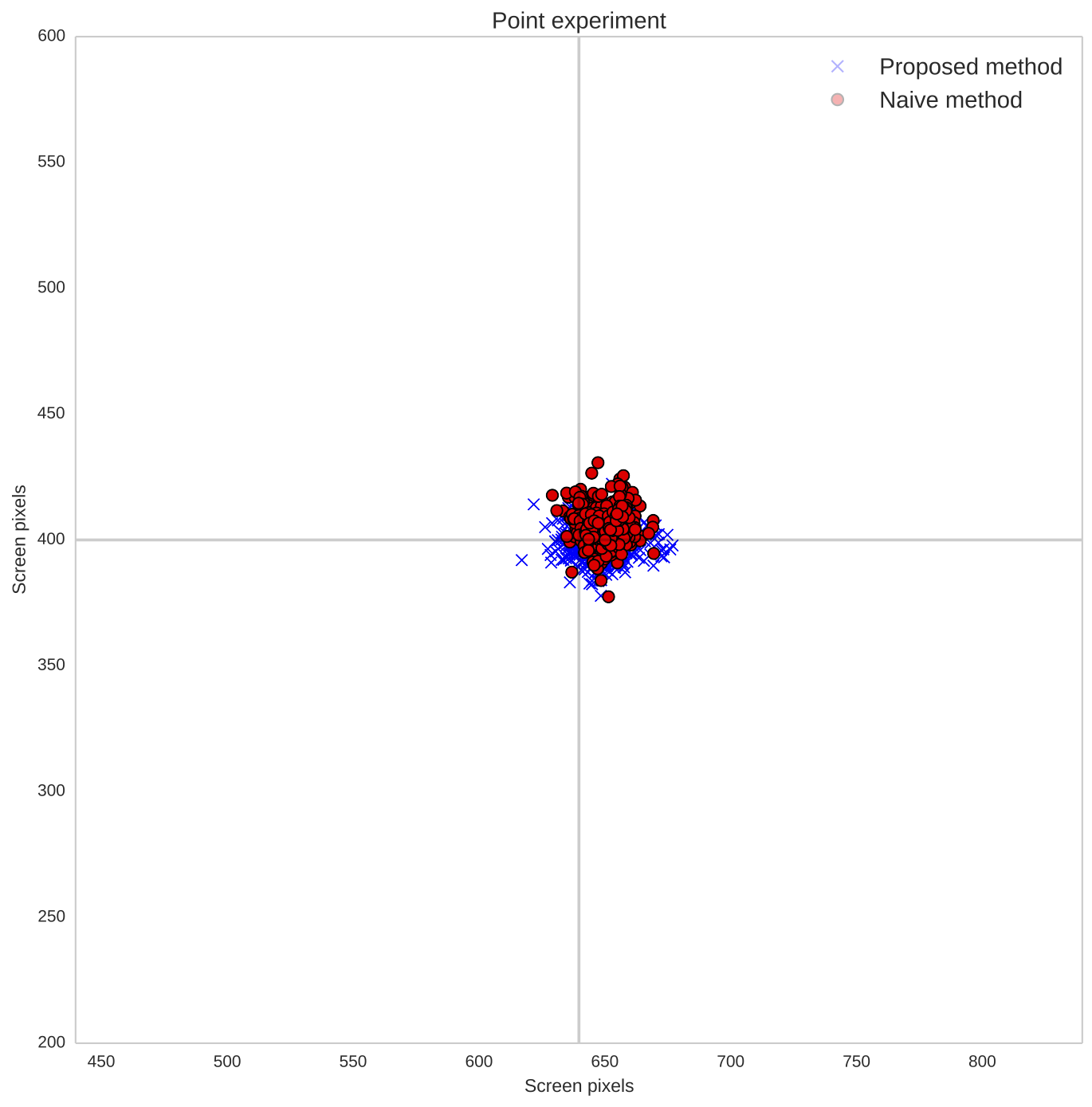} &%
   \includegraphics[width=.25\textwidth]{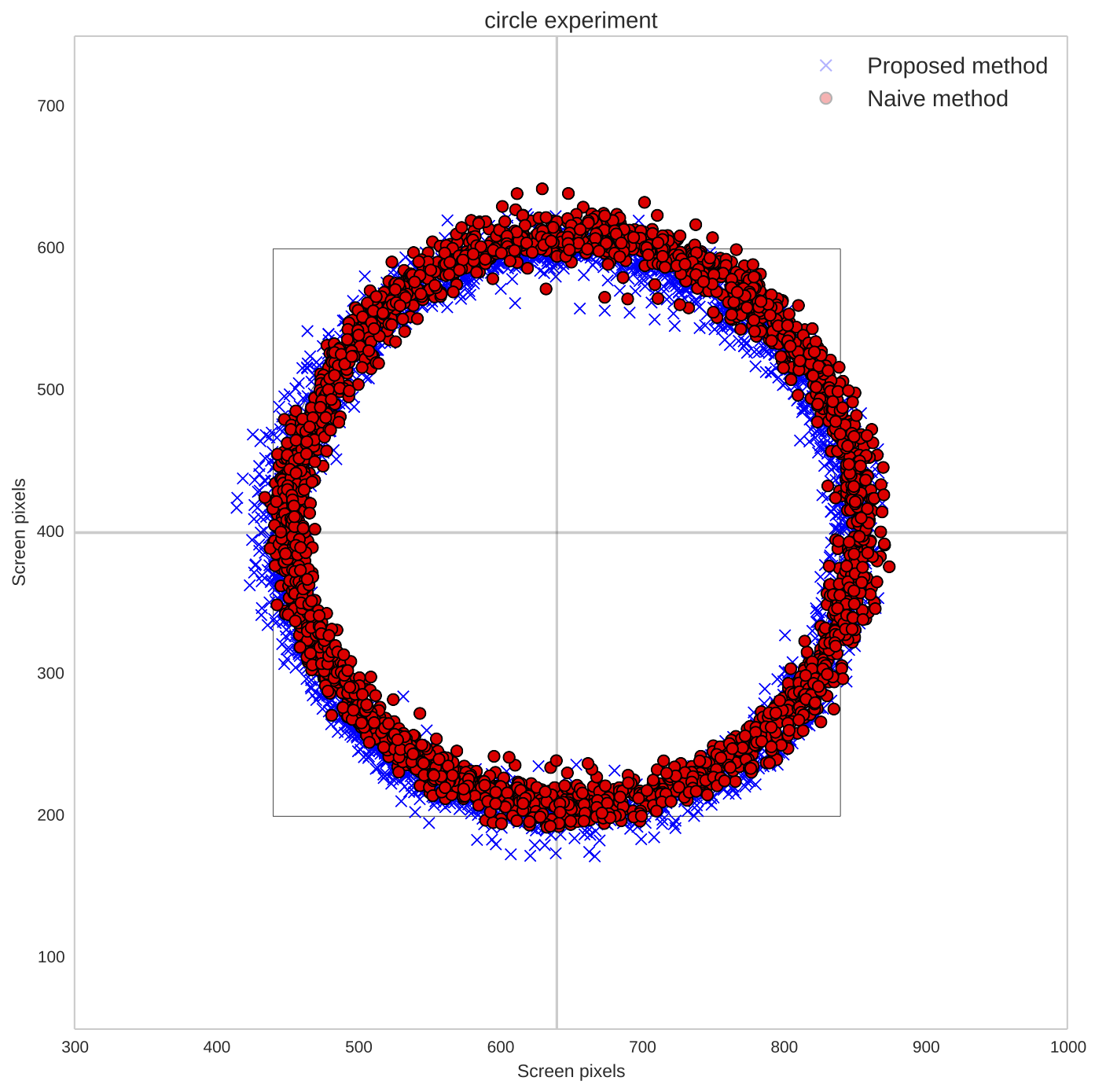}
 \end{tabular}
    \caption{Blue cross represents our method, while red point represents naive method. Tasks from the left: horizontal line, point and circle task.}
    \label{fig:task_results}
\end{figure*}

\begin{figure}[h!]
\centering
\includegraphics[width=\linewidth]{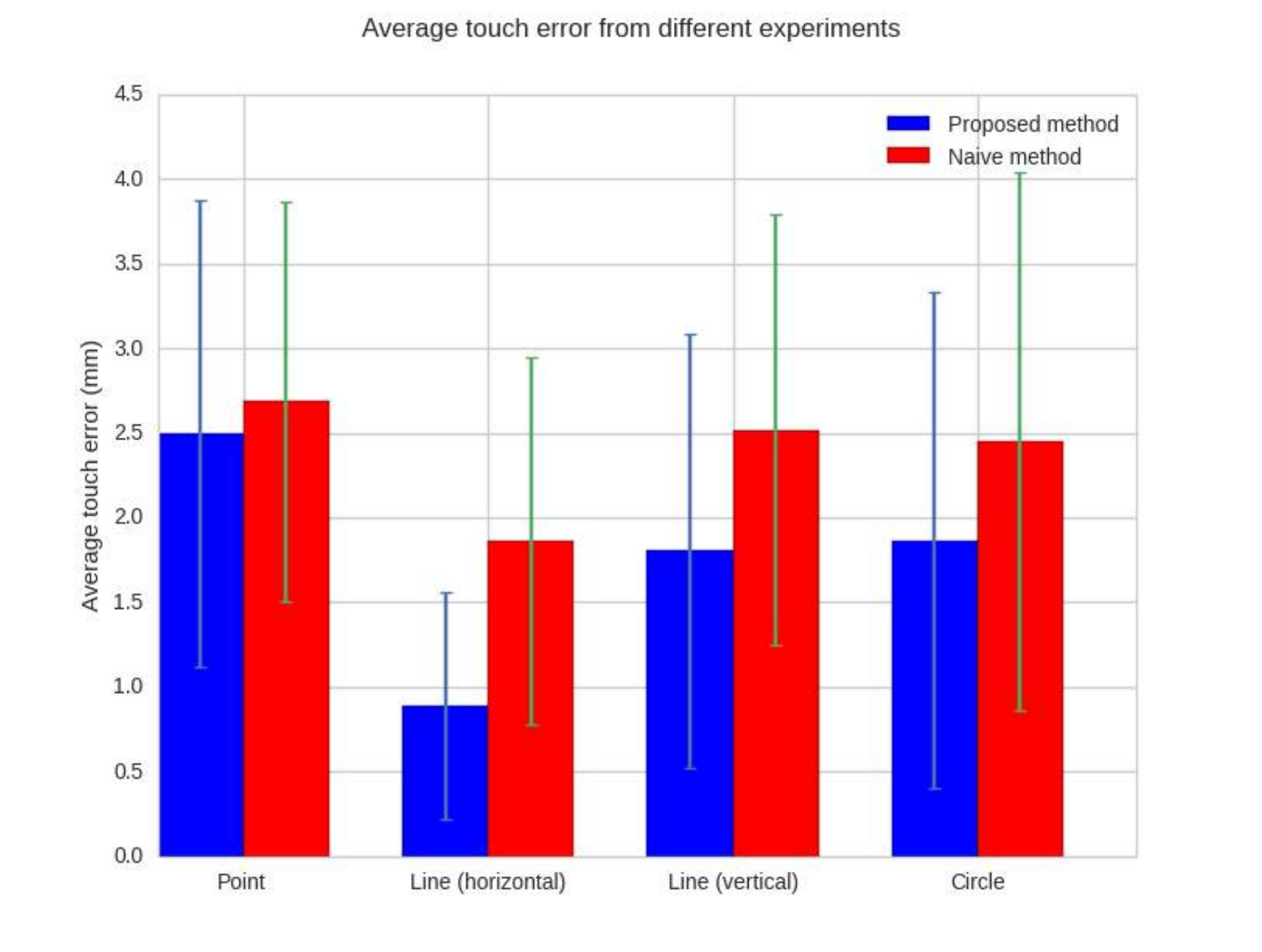}
\caption{Average touch errors in mm for 4 tasks.}
\label{fig:task_error}
\end{figure}

Each user was asked to touch the center of the cross for 20 times repeatedly (see Fig.\ref{fig:tasks}(a)).  We record the results from our method as well as the naive results simultaneously for each single touch(see Fig.\ref{fig:task_results}(a)). For an error metric we measure the L2 distance from the known center of cross to detected results. Users are also asked to track a horizontal line, a vertical line and a circle (see Fig.\ref{fig:tasks}(b, c, d)). Similarly, we get both results from our method and naive method (see Fig.\ref{fig:task_results}(b, c, d)) and calculate the shortest L2 distance between the prediction and reference. The average touch errors in millimeter for all 4 tasks are presented in Fig.\ref{fig:task_error}. Our method outperforms the \cite{Wilson:2010} method in terms of average accuracy. 

\subsection{Finger Disambiguation}

For finger designation we did not perform comparative analysis, instead we hereby report on the accuracy and robustness of the algorithm in the cases of single-touch hand poses and the more challenging multi-touch hand poses.

\subsubsection{Single Touch} \label{singleTouch}

10 volunteers were invited for the single-touch detection evaluation. We examine the following two aspects:

\begin{enumerate}
  \item Successful detection of finger-touch presence.
  \item Correct identification of the intended finger.
\end{enumerate}

Each participant was asked to complete the following task: touch the tabletop surface with each finger (in the order of thumb, index, middle, ring and little finger) of both hands repeatably till instructed to stop, and count how many times their finger touched the surface. The system automatically identifies the finger and stops after 20 touches from each finger were detected.

For each finger, we define the average number of actual touches, acquired directly from the participants, $n$ ($n\geq 20$) as ground truth, and the ratio of correctly identified touches $c/20$ (true positives) to be the accuracy metric, where $c$ ($c\leq 20$) is the number of correctly identified touches. Higher value of $n$ implies larger difficulty of the system detecting fingers.

Participants in the above experiments could touch the surface with any hand gesture at will. A supplementary experiment is conducted further to examine the accuracy of finger disambiguation when the subjects are instructed to use gestures with reasonable finger angles w.r.t the table.

\begin{figure}[h!]
\centering
\includegraphics[width=\linewidth]{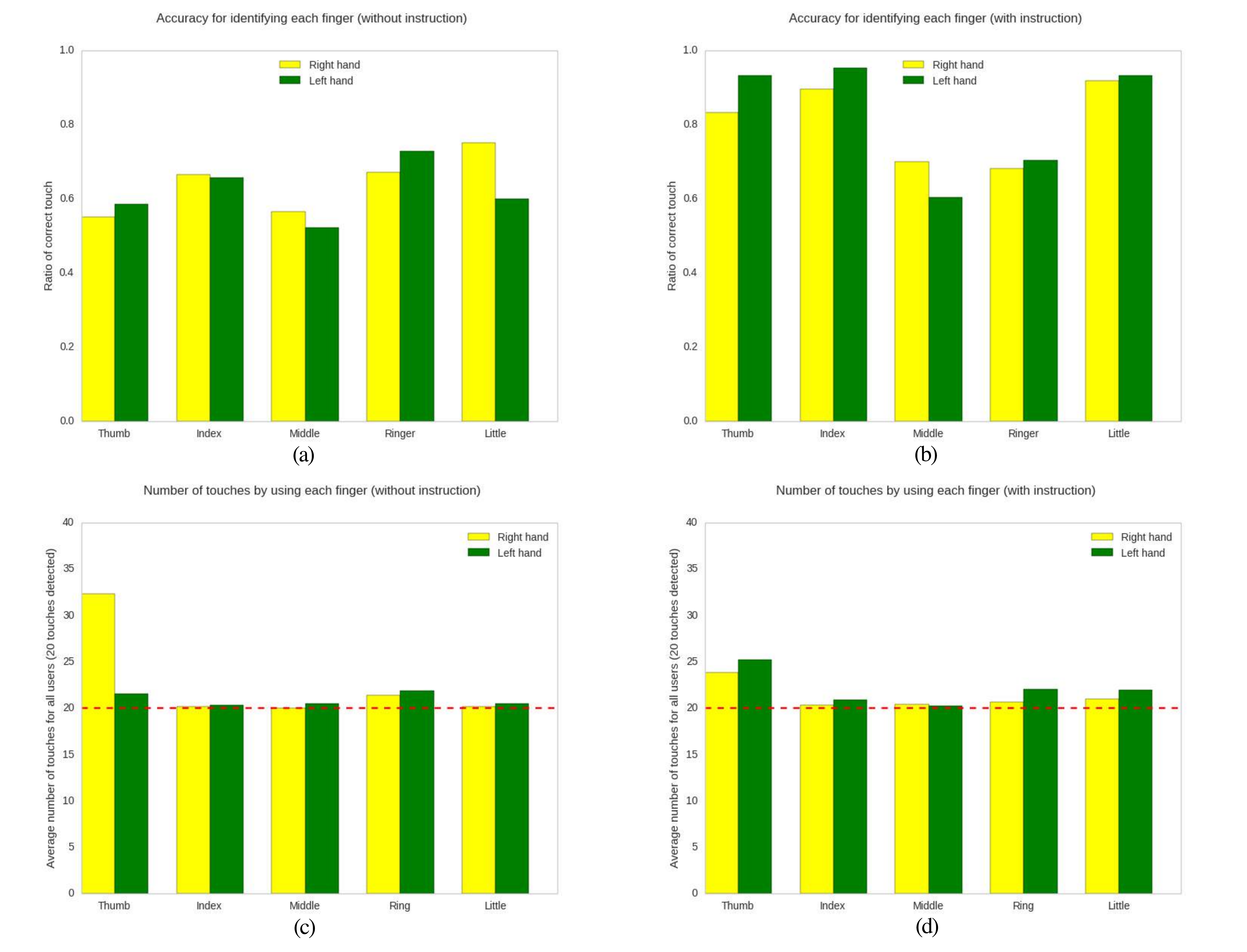}
\caption{The first column corresponds to hand poses without instruction while the second column corresponds to poses with instruction. (a, b) average ratio of accurately identified touches. The confusion matrices show the system's confusion w.r.t finger designation.}
\label{fig:single_results}
\end{figure}

Our data shows that without instruction the thumb and middle finger create more false-negatives than the other fingers, which suggests a larger difficulty for our system to detect them accurately. This can possibly be attributed to a phenomenon we observed in the course of experiments that subjects generally touch with the side of their thumb instead of the pad.

After instruction, thumb, index, and little fingers are found to be relatively well detectable with detection ratios respectively. Our system occasionally confuses the middle finger with the index finger owing to their resemblance in certain gestures.

The difference between left and right hand in terms of both accuracy and presence metric is inconsequential. Moreover, the system performs consistently across hands from different users.

\subsubsection{Multi-touch}
\begin{figure}[h!]
\centering
\includegraphics[width=\linewidth, scale=0.8]{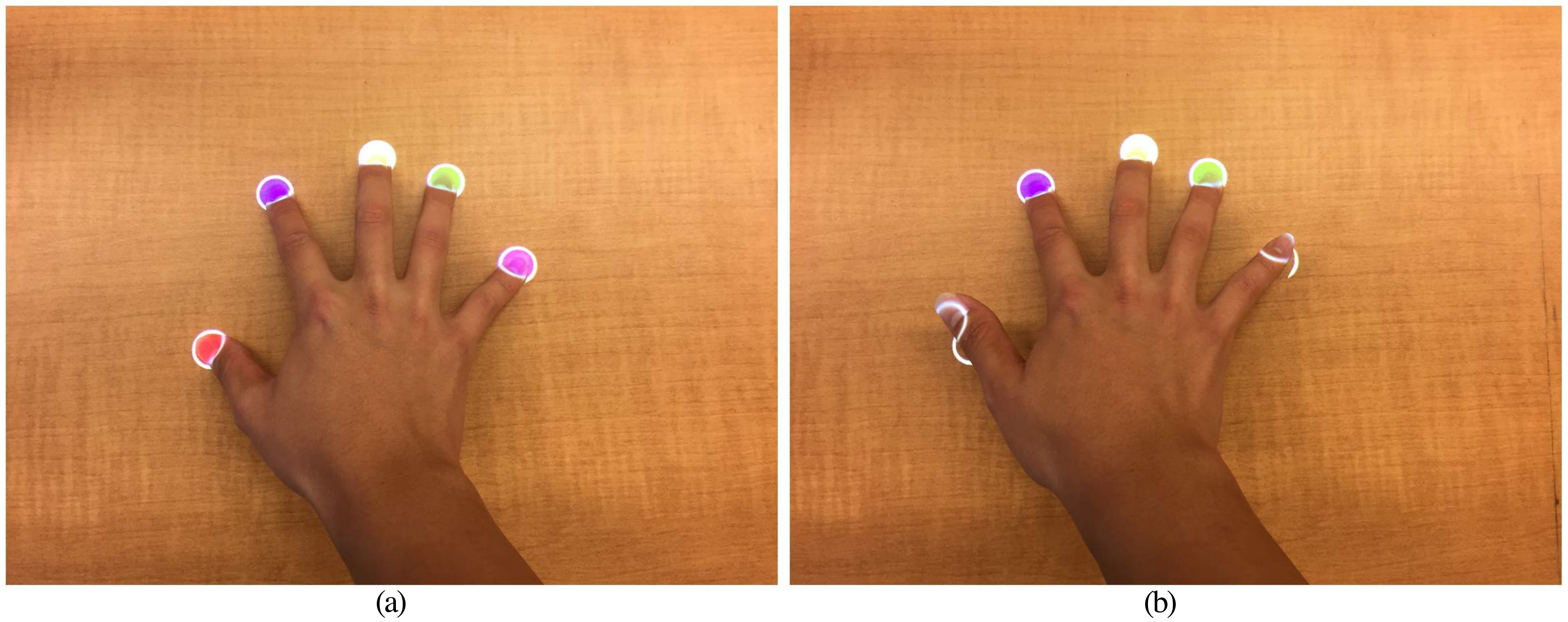}
\caption{(a) cover all 5 circles with corresponding fingers. (b) cover index, middle and ring circles with corresponding fingers.}
\label{fig:multi_task}
\end{figure}

In the multi-touch experiments all participants had instruction on the gesture to use for touching, see Fig.\ref{fig:multi_task}. The following analysis is based on outcomes from right hand alone, as no noticeable difference could be observed between left and right hands.

\begin{figure}[h!]
\centering
\includegraphics[width=\linewidth]{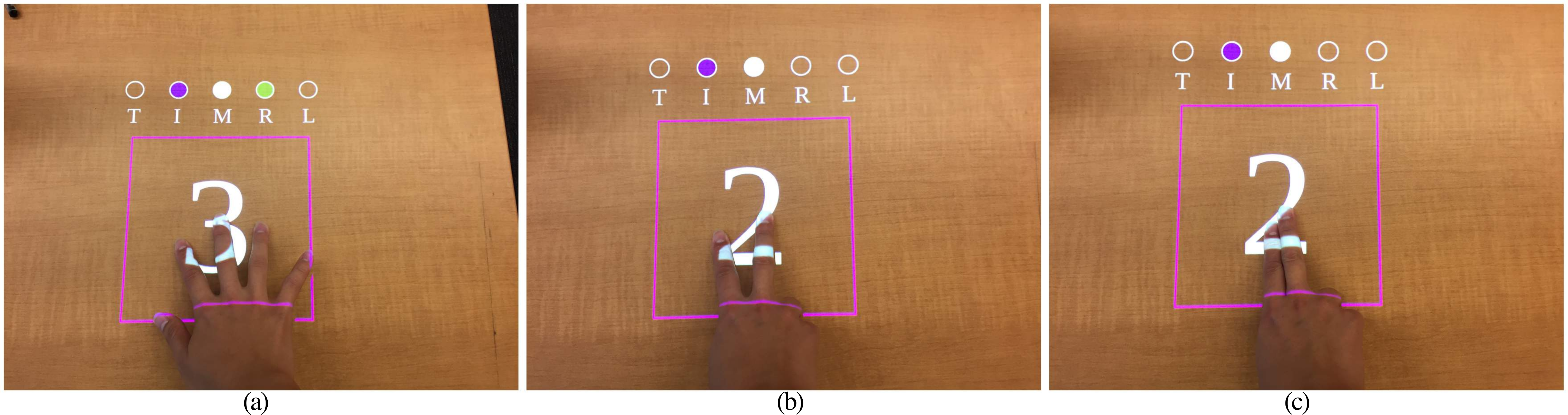}
\caption{The system computes the number of fingers as well as finger identifications. (a) touch with index, middle and ring fingers. (b) touch with index and middle fingers. (c) touch with index and middle fingers together.}
\label{fig:multi_touch}
\end{figure}

For multi-touch accuracy determination, we instructed the users to place their finger pads on the colored dots projected onto the surface, as seen in Fig.\ref{fig:multi_task}. We count the number of times our system mistakenly assigned wrong finger labels. As an example, even though the presence of index and middle finger can be sensed correctly, the system might still label index finger as middle and middle finger as index.

We asked our participants to perform the touches 10 times each with: 2 fingers (index and middle), 3 fingers (index, middle and ring), 4 fingers (index, middle, ring and little), and all 5 fingers and to make a best effort to cover the projected circles with the corresponding fingers.

\begin{figure}[h!]
\centering
\includegraphics[width=\linewidth]{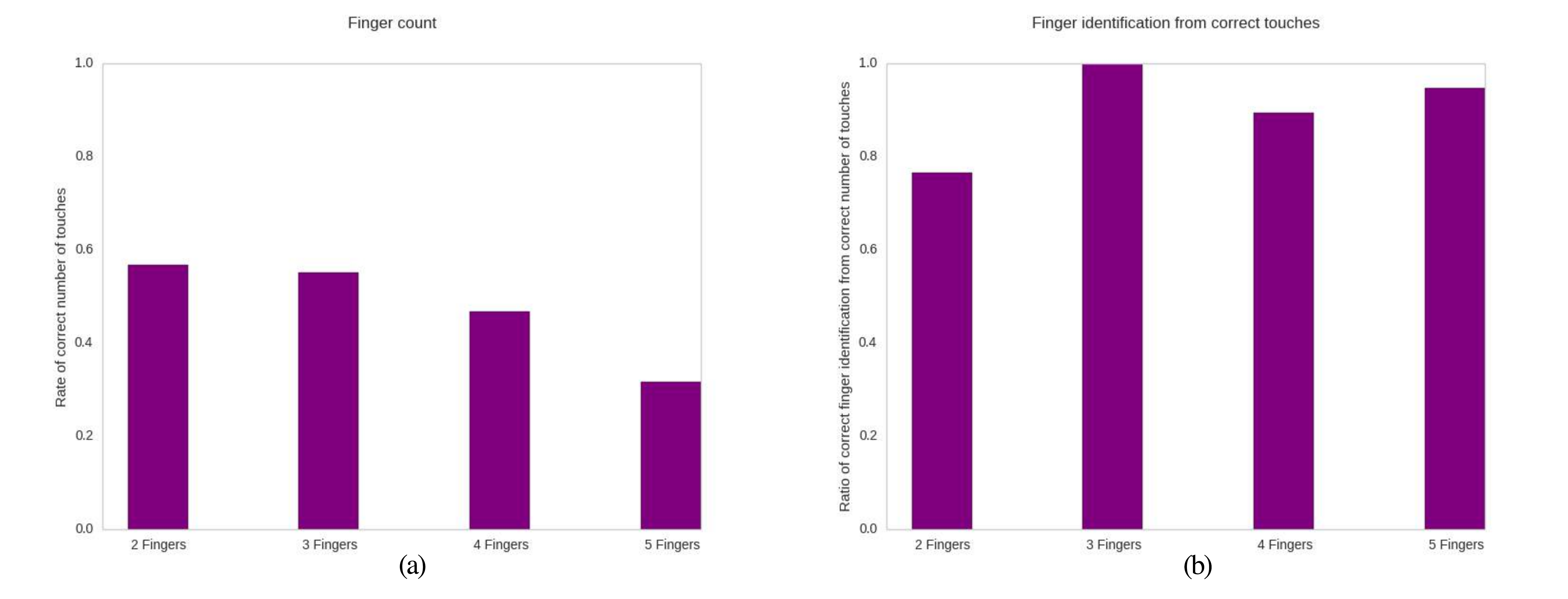}
\caption{(a) how many multi-touches include correct number of fingers. (b) how many multi-touches are accurately identified.}
\label{fig:multi_results}
\end{figure}

As a metric for accuracy in the multi-touch task we count the number of times our system was able to detect all the intended figures with a correct \textit{total designation} (the right label for the right finger) for each condition: 3, 4 and 5 fingers. The results (illustrated in Fig.\ref{fig:multi_results}) show that our system struggles with \textit{total designation}, especially in the 5 finger case where it only detects fully at a third of the trials. However, we observed that participants had a tendency to not touch all the fingers at once but one at a time, which confused our automatic total designation measure.

We designed another metric for finger designation accuracy: $b/a$, where $b$ stands for the number of \textit{total designations} and $a$ is the number of \textit{partial designations} (where the right number of fingers is detected but the label designation is not true). Fig.\ref{fig:multi_results}(b) shows statistics on the accuracy metric. We conclude that our system can distinguish different fingers effectively as long as correct number of fingers is detected.

\section{Applications}
We devised several applications for using the new finger designation features of our method.

\subsection{Virtual Keyboard}
\begin{figure}[h!]
\centering
\includegraphics[width=\linewidth]{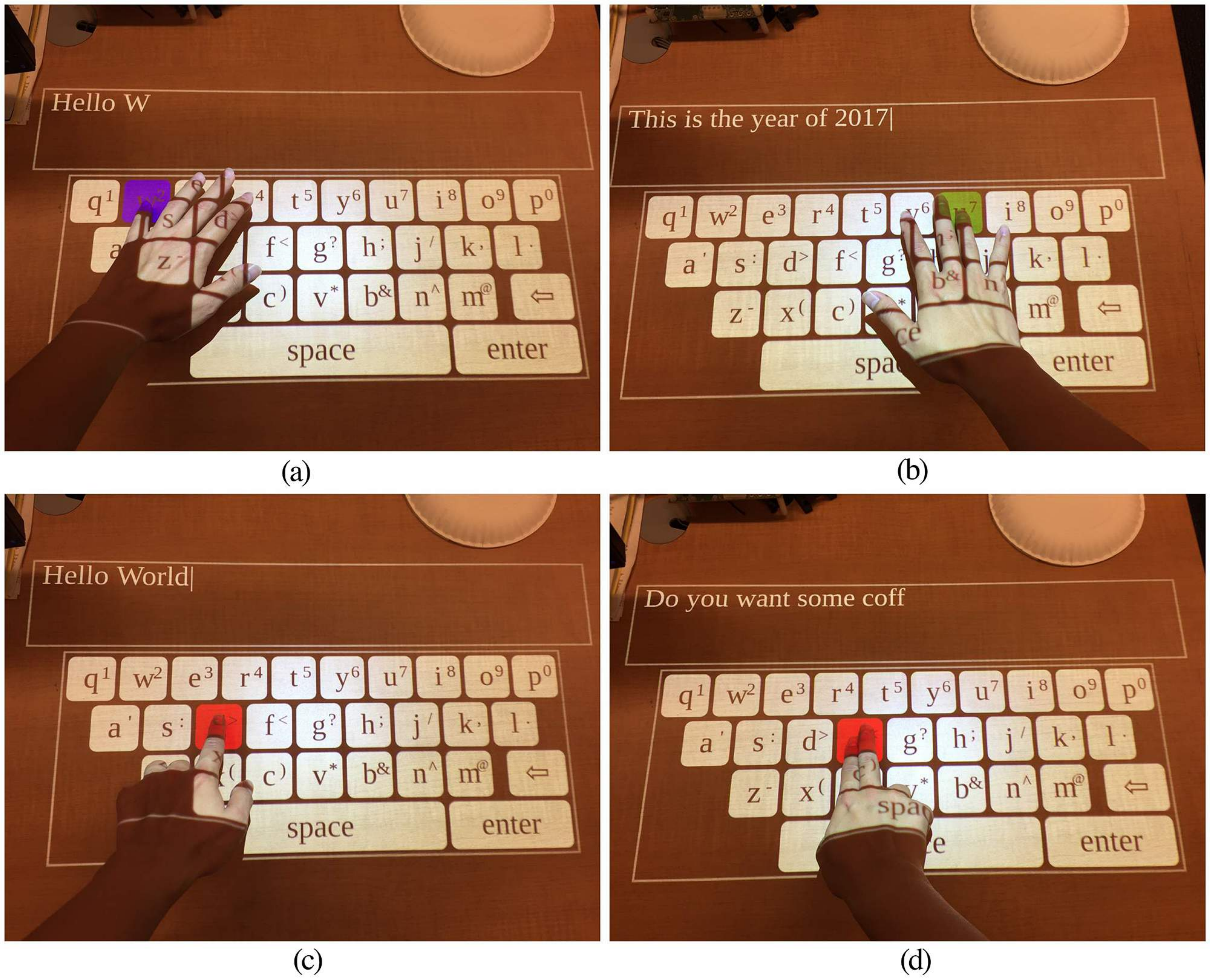}
\caption{Our novel keyboard usages. (a) type upper case letters with little fingers from both hands. (b) type numbers and non-alphanumeric characters with ring fingers from both hands. (c) type lower case letters with index finger, middle finger and thumb from both hands. (d) type double letters with two fingers.}
\label{fig:keyboard}
\end{figure}

On regular keyboards, most non-alphanumeric characters can only be typed using a combination of multiple keys (usually ``shift'' plus other keys). ~\cite{Gupta:2016} proposed a text entry technology that relies on finger identification to get rid of the need for a ``shift''. However, they can only identify index finger and middle finger. On the other hand, users need to wear specialized gear to enable the identification feature. In this work, we present an enhanced text entry on a projected keyboard (see Fig.\ref{fig:keyboard}) which utilizes finger identification.

On this keyboard, thumb, index and middle finger are only allowed to type lowercase letters (see Fig.\ref{fig:keyboard}(c)). Little finger is reserved for typing uppercase letters (see Fig.\ref{fig:keyboard}(a)). The ring finger is responsible for typing numbers and non-alphanumeric characters (see Fig.\ref{fig:keyboard}(b)). Users can even type double letters like in Fig.\ref{fig:keyboard}(d) by touching with two fingers pressed together (e.g. to get ``ff'' when typing once with the index+middle fingers on the ``f'' key).

The same 10 volunteers were used to evaluate the keyboard by typing 20 samples from a small but representative dataset of English~\cite{Yi:2017}. We utilize WPM (Words Per Minutes) and CER (Character Error Rate) to evaluate the responding speed and the error rate of our keyboard. Among all test takers, we measured the highest WPM at 20.70 and the lowest WPM at 8.39, the smallest CER as 6.36 and the biggest CER as 41.77. The average WPM and CER are 12.72 and 21.63.

By adding a auto completion, we increase the speed of our enhanced keyboard by 57\%. We asked the volunteers to complete the same task, which resulted in a better WPM of 19.98.

\begin{figure}[h!]
\centering
\includegraphics[width=\linewidth]{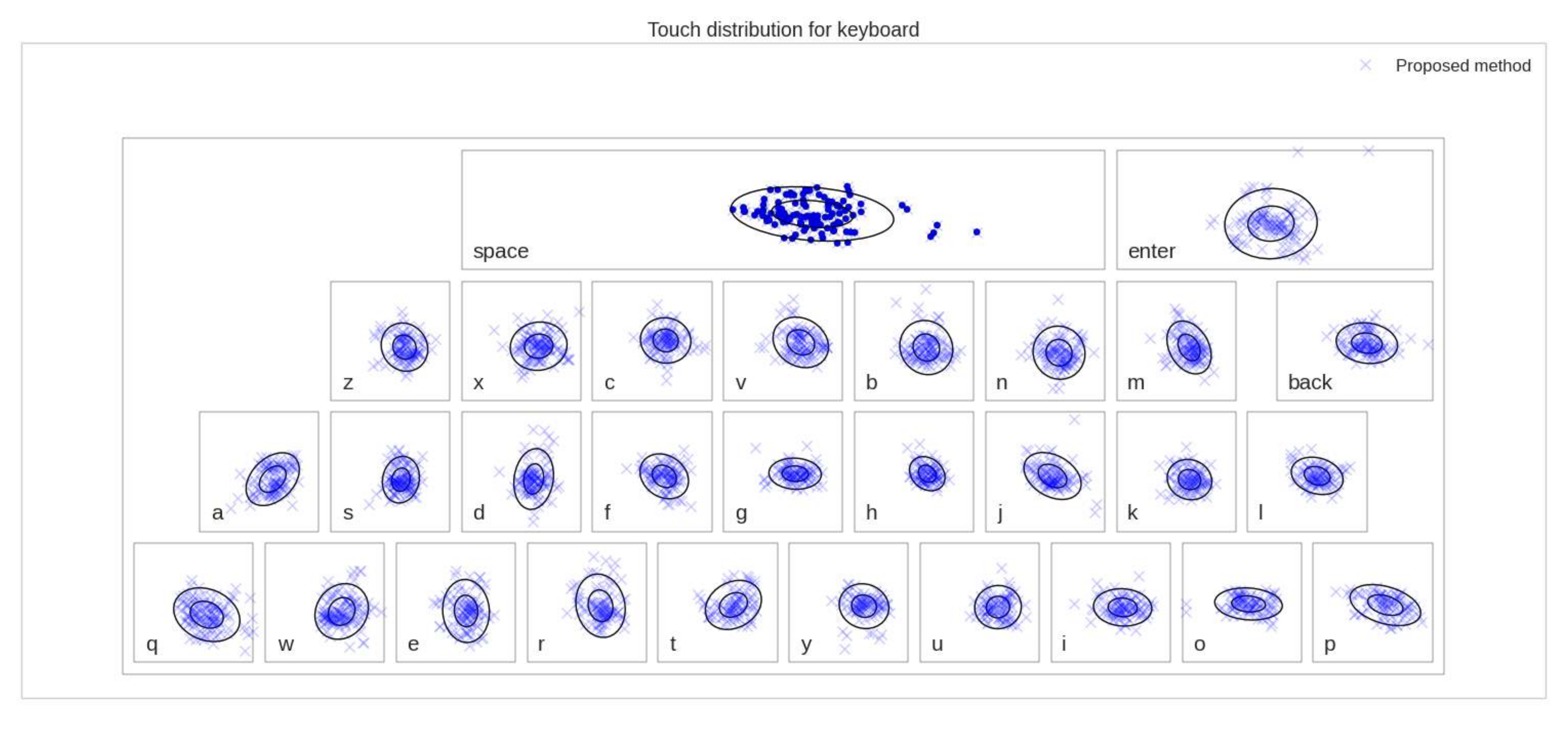}
\caption{The distribution of touches on our enhanced keyboard.}
\label{fig:keyboard_dis}
\end{figure}

We record the position of all touches on our enhanced keyboard collected from 10 volunteers, and show the distribution in Fig.\ref{fig:keyboard_dis}. For each key, we plot two confidence ellipses corresponding to 1 standard error and 2 standard error.

\subsection{Other Potential Applications} \label{potentialApp}
Our method can also be used in many other potential applications: Drawing board and e-books.

\begin{figure}[h!]
\centering
\includegraphics[width=\linewidth]{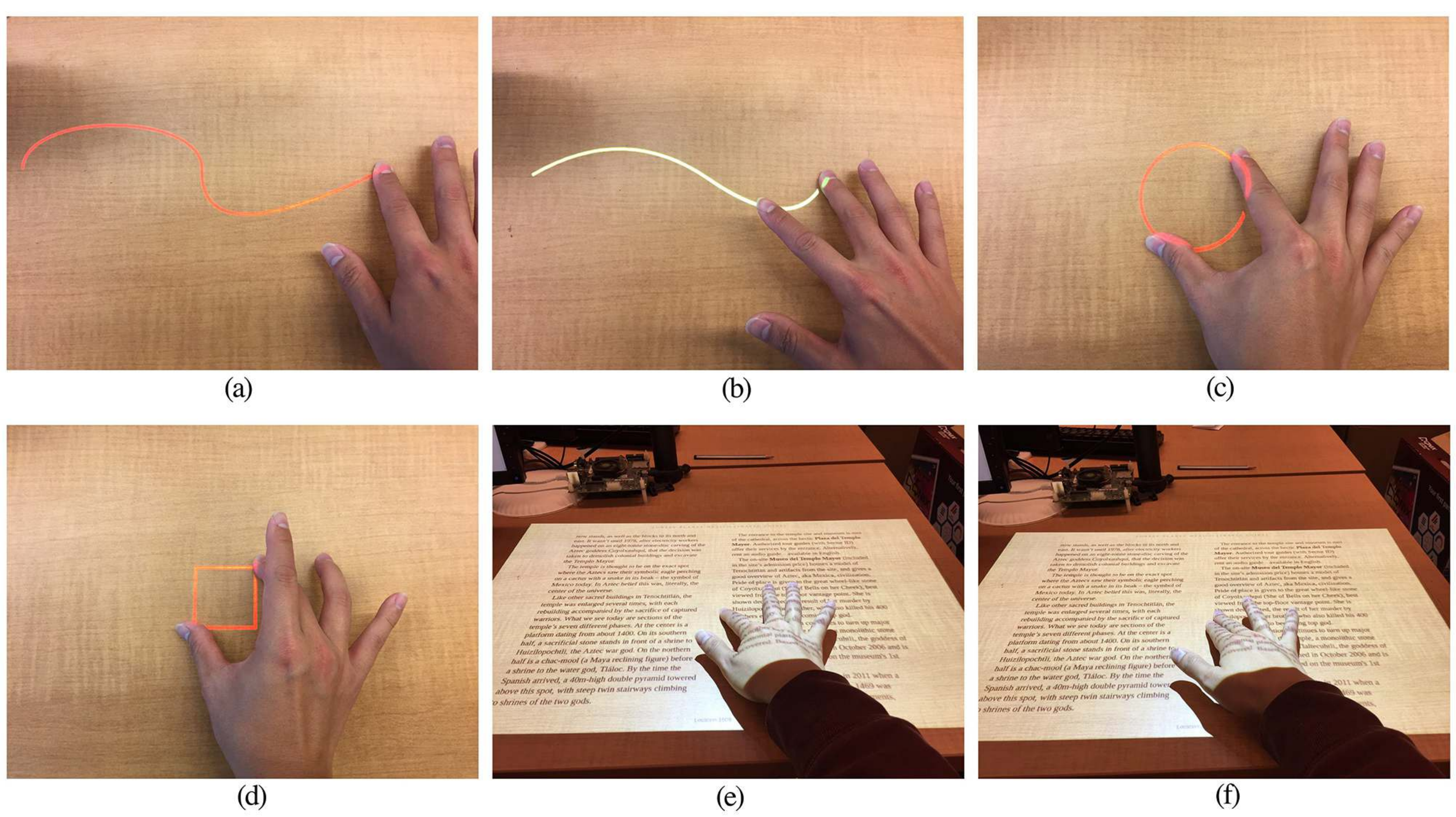}
\caption{(a) draw a red curve with index finger, (b) draw a yellow curve with middle finger, (c) draw a circle by using thumb and index finger together, (d) draw a square by using thumb and middle finger together, (e) turn one page by using one finger, (f) turn two pages by using two fingers}
\label{fig:app2}
\end{figure}

In Drawing board, users can draw different shapes by using different combinations of gestures (see Fig.\ref{fig:app2}(c, d)). Also they can draw lines and curves in different colors by using different fingers (see Fig.\ref{fig:app2}(a, b)). By using their thumb only, users are allowed to erase the existing content on the display surface.

Also, our method can be used in electronic books. In current e-books, people have to read the book page by page, which is tedious. By using our pose-aware touch detection, users can decide the number of pages to be turned by sliding with different number of fingers (see Fig.\ref{fig:app2}(e, f)).

\section{Limitations and Future Work}
There are several limitations of this work, which we see as opportunities of future work. We find hand pose estimation becomes the bottleneck in our working system, improve the post estimation can greatly boost the accuracy and stability of the whole system. 

First, as the most complicated and flexible part of the human body, hand needs to be represented by a large parameter space due to the high degree of freedom (DOF). An excess of parameters leads to unstable predicted results in pose estimation. As mentioned in~\cite{OberwegerWL15}, adding PCA on the top of CNN helps reducing the parameters in the network and can thus providing more stable results.

Second, occlusion remains a big problem in hand pose estimation due to hardware limitation. Using multiple cameras with different viewpoints can efficiently alleviate the problem. However, it makes the physical setup much more complicated. Adding multi-view information~\cite{GeLYT16} is an alternative to mitigate the effect of occlusion.

\section{Conclusion}
Our touch detection system introduced novel capabilities to standard touchable projection with structured-light sensors: finger designation, touch-blobs decomposition, and estimated touch points based on a hand skeleton model that is more robust to noise. These additional features may inform more deliberate touch gestures, such as pinch-to-zoom and multi-finger tap or drag.

We evaluate our proposed method in different aspects and prove the accuracy and robustness of the system. We also proposed a novel virtual keyboard which use finger identification as a new feature.

\section{Acknowledgements}
We would like to thank Naveen Rai, Fan Wang and Shahira Abousamra for their help in producing this paper.
We thank Nvidia for their generous gift of a Titan Xp GPU that we used towards training and running our algorithms.
We thank Tulip Interfaces for their generous donation of a projector and camera that we used in our setup.

% BALANCE COLUMNS
\balance{}

% REFERENCES FORMAT
% References must be the same font size as other body text.
\bibliographystyle{SIGCHI-Reference-Format}
\bibliography{sample}

\end{document}